\documentclass[aps,pre,twocolumn,superscriptaddress]{revtex4-1}

\usepackage{amsmath}
\usepackage{graphicx}
\usepackage{mwe}
\usepackage[caption=false]{subfig}
\usepackage[usenames, dvipsnames]{color}
\usepackage{verbatim}

\begin{document}


\title{Compression stiffening in biological tissues: on the possibility \\of classic elasticity origins}


\author{T. A. Engstrom}
\email[]{taengstr@syr.edu}
\affiliation{Department of Physics, Syracuse University, Syracuse, NY 13244, USA}
\author{K. Pogoda}
\altaffiliation{Institute of Nuclear Physics, Polish Academy of Sciences -- PL-31342, Krakow, Poland}
\affiliation{Institute for Medicine and Engineering, The University of Pennsylvania -- Philadelphia, PA 19104, USA}
\author{K. Cruz}
\affiliation{Institute for Medicine and Engineering, The University of Pennsylvania -- Philadelphia, PA 19104, USA}
\author{P. A. Janmey}
\email[]{janmey@mail.med.upenn.edu}
\altaffiliation{Departments of Physiology and Physics \& Astronomy, The University of Pennsylvania -- Philadelphia, PA 19104, USA}
\affiliation{Institute for Medicine and Engineering, The University of Pennsylvania -- Philadelphia, PA 19104, USA}
\author{J. M. Schwarz}
\email[]{jschwarz@physics.syr.edu}
\affiliation{Department of Physics, Syracuse University, Syracuse, NY 13244, USA}


\date{\today}

\begin{abstract}
Compression
  stiffening, or an increase in shear modulus with increasing
  compressive strain, has been observed in recent rheometry
  experiments on brain, liver, and fat tissues.  Here, we extend the
  known types of biomaterials exhibiting this phenomenon to include
  agarose gel and fruit flesh. Further, we show that two different
  results from classic elasticity theory can account for the
  phenomenon of linear compression stiffening. One result is due to
  Barron and Klein, extended here to the relevant geometry and
  pre-stresses; the other is due to Birch. For incompressible
  materials, there are no adjustable parameters in either
  theory. Which one applies to a given situation is a matter of
  reference state, suggesting that the reference state is determined
  by the tendency of the material to develop, or not develop, axial
  stress (in excess of the applied pre-stress) when subjected to
  torsion at constant axial strain. Our experiments and analysis
  also strengthen the notion that seemingly distinct animal and plant
  tissues can have mechanically similar behavior under certain conditions.
\end{abstract}


\maketitle


\section{Introduction}

The effect of pre-stress on a biological tissue's elastic moduli and related sound velocities, etc. is an interesting question, given that a living tissue confined in volume generically develops pre-stress in the form of homeostatic pressure. This condition is characterized by a steady-state of cell division and death processes~\cite{Basan09}. For vascularized tissue, an upper limit on homeostatic pressure is set by the $\sim10$~kPa blood pressure. \emph{Ex vivo} shear stiffness of mammalian brain matter is $\sim1$~kPa, for comparison~\cite{Weickenmeier16, Budday17}. One might naturally ask how, or whether, the latter value would be different in the case of living or otherwise pre-stressed tissue. Very recent
results using magnetic resonance elastography indicate the shear modulus of living brain tissue increases linearly with intracranial (homeostatic) pressure~\cite{Arani18}.

A series of recent parallel-plate rheometry experiments have explored pre-stress
effects in animal tissue and biopolymer network samples of
characteristic size $\sim 1$~cm, by subjecting them to a combination
of static axial compression and $\sim 1$~Hz torsional
oscillations~\cite{Pogoda14, Perepelyuk16, Licup15, Vahabi16,
  vanOosten16, Mihai15}. To avoid slippage during torsion, and also to
facilitate axial tension, adhesive contact is typically made between
the rheometer plates and the ends of the cylindrical
sample~\footnote{Adhesive contact also provides a means of reducing
  stresses generated by the sample's own weight; for 5 mm cubes of
  brain tissue, the associated gravitational strains can apparently be
  as large as 60\%~\cite{Budday17}, an order of magnitude larger than
  one expects based on a simple calculation.}. It has been pointed out
that such adhesive boundary conditions effectively constrain the
lateral dimensions of a sufficiently thin sample, resulting in a
volume change when axial force is applied~\cite{Vahabi16, vanOosten16}.
In this thin film limit, one expects stresses within a
fluid-containing tissue sample are redistributed into a state of near
hydrostatic pressure. Thus, by adjusting the sample geometry and/or
boundary conditions, parallel-plate rheometers provide a convenient
way to measure the effect of various states of pre-stress on the shear
storage and loss moduli of tissues. 

Determining how pre-stress affects moduli may be relevant for
diagnosing diseases associated with tumors or other abnormal growth,
and identifying their regions of extent. Recent work, for example, has
addressed the shear response of brain tissue (both normal and that
isolated from human glioma tumors), as a function of pre-stress levels
expected \emph{in vivo} from homeostatic pressure considerations as
well as increased vascularization of the tumors~\cite{Pogoda14}.
Similar measurements were carried out on liver tissue (both normal and
that affected by fibrosis)~\cite{Perepelyuk16}. In all four of these
cases, shear storage modulus is reported to increase with applied
uniaxial compression. The authors refer to this phenomenon as
compression stiffening. Meanwhile, when essentially the same
experiment is done with the biopolymer network materials collagen and
fibrin (major components of the extracellular matrix), the opposite
effect is found: shear storage modulus decreases with uniaxial
compression, but increases with extension~\cite{Licup15, Vahabi16}. One
potentially unifying feature of these diffferent results, however, is
an observed linear relationship between shear storage modulus and
uniaxial pre-stress~\cite{Perepelyuk16,Vahabi16}.

Compressional stiffening has now been observed in a number of
different animal tissues; might it also be a property of plant tissue?
Individual plant cells contain cell walls, vacuoles, and chloroplasts,
which individual animal cells do not.  Plant cell walls allow the
cells to withstand turgor pressures on the scale of
megaPascals~\cite{Schopfer06}, and presumably result in plant tissue
typically having larger storage moduli than animal tissue at the many
cell scale. And while plant tissue has long been modeled as an elastic
solid~\cite{Ghysels09}, as has animal tissue, it would be interesting
to quantitatively compare the two at both small and large strain.

To provide an interpretation for the observed compressional stiffening, we point
out a subtlety concerning the measurement of elastic constants of a
material under pre-stress, and argue that certain instances of linear
compression stiffening can be explained by properly accounting for
pre-stress in the rheometry experiments. The theory involved was developed in the context of condensed matter at high pressure~\cite{Grimvall12}, and has not commonly been applied to soft matter at physiologically relevant pressures. While pre-stresses in soft matter systems may be small in
absolute terms, they can be large in comparison to the elastic moduli,
as already mentioned. Our pre-stress calculation emphasizes the role
of boundary conditions in determining whether the applied uniaxial
stress remains uniaxial within the sample, or is redistributed into an
isotropic stress. Meanwhile, other instances of linear compression
stiffening are not readily explained by the pre-stress
theory. Instead, they are consistent with a conceptually different
theory in which hydrostatic compression and shear are superposed on a
zero-stress reference state. Thus the present work argues that {\it both} theories are applicable to linear compression stiffening; which one works in a given situation depends upon the nature of the reference state. 

\section{Experiments: materials and methods}

{\it Animal tissue samples:} 
The dependence of shear modulus on compressive strain has previously been reported for mouse brain~\cite{Pogoda14}, liver~\cite{Perepelyuk16}, and fat~\cite{Mihai15}, as has the relation between axial stress and axial strain.   Here we replot these data to show shear modulus as a function of axial stress.   Details of the sample preparation and rheological methods are provided in Refs.~\cite{ Pogoda14,Perepelyuk16,Mihai15}. Briefly, animal tissues were cut into disk-shaped samples using an 8 mm diameter stainless steel punch. Fibrin gel, with a shear modulus greater than that of the tissues, was used to glue the sample to the rheometer plate, and a normal force of 1 g was applied to ensure contact between the top of the sample and the upper plate.  This state was assumed to approximate the zero stress state.  The shear modulus of the samples was measured on a strain-controlled Rheometrics fluids spectrometer III (Rheometrics, Piscataway, NJ), which can also measure normal forces simultaneously with torque. Axial strain was applied by changing the distance between the parallel plates, and the resulting axial stress was measured 30 s after changing the gap.

{\it Mango samples:} 
Mango fruit flesh was obtained from a ripe mango.  Samples were cut into 10 mm high and 20 mm wide disks, with the long axis parallel to the seed orientation, using a 20 mm tissue punch.  All samples were from the same fruit. The shear elastic and viscous modulus of mango flesh was measured using a Malvern Kinexus lab+ rheometer and rSpace software (Westborough, MA) using a 20 mm parallel plate geometry.  Because mango is slippery, 20 mm sandpaper discs were used to ensure contact between the plates and the sample. Shear modulus was measured at 1 rad/sec and 5\% oscillatory strain.  Triplet mango samples were measured under increasing compression, up to 20\% of the original sample height, in steps of 4\% each.

{\it Agarose gel samples:}
2\% agarose solution was prepared by dispersing the appropriate amount
of polymer in distilled water at 100 deg C, while stirring until
complete dissolution. Agarose gels were prepared by pouring the above
solution into a mold and allowing the gelation for 24 h at room
temperature. Samples were cut from one big chunk into 9.5, 3.8 and 1.8
mm high and 20 mm wide disks using a 20 mm tissue punch. Rheology
measurements were performed using Malvern Kinexus lab+ rheometer with
a 20 mm parallel plate geometry. Shear modulus measurements were carried out at a
frequency of 1 Hz, shear strain amplitude of 2\%, and axial compressive strain increasing in 
steps of 5\% up to maximum value 25\%.

\section{Barron-Klein approach and its application to parallel plate rheometry}

\subsection{Background}

A 1965 paper by Barron and Klein (hereafter, BK1) treats rigorously the problem of calculating elastic constants of a solid under pre-stress~\cite{Barron65}. Taylor expanding the energy density around the pre-stressed reference configuration yields
\begin{equation}
\frac{\Delta U}{V} = S_{ij}u_{ij} + \frac{1}{2}Q_{ijkl}u_{ij}u_{kl} + \dots, \label{energy_density}
\end{equation}
where $u_{ij}$ is the combined deformation due to the pre-stress $S_{ij}$, and any other stresses subsequently applied to the reference state. The key point (made 15 years prior to BK1) is that the presence of the linear term modifies the symmetry properties of $Q_{ijkl}$ from those of the usual rank-four elastic modulus tensor~\cite{Huang50}. In particular, invariance of the energy density under a rigid rotation requires that
\begin{eqnarray}
Q_{ijkl}-Q_{jikl} = S_{jl}\delta_{ik} - S_{il}\delta_{jk},\\
Q_{ijkl}-Q_{ijlk} = S_{jl}\delta_{ik} - S_{jk}\delta_{il},\\
Q_{ijkl}-Q_{jilk} = S_{jl}\delta_{ik} - S_{ik}\delta_{jl}.
\end{eqnarray}
However, BK1 shows that there is a tensor $c_{ijkl}$, that in the special case of isotropic pre-stress $S_{ij}=-P\delta_{ij}$ where $P$ is pressure, inherits all the symmetries of the usual elastic modulus tensor, and enters the stress-strain relationship and equation of motion in the usual way. (Homogeneous deformation is assumed in their analysis.) The cost of this finite pressure generalization of the zero-stress elastic constants is that $c_{ijkl}$ is not equal to the second derivative of energy density with respect to strain, rather that plus a ``pressure correction'' term:
\begin{equation}
c_{ijkl} = \frac{1}{V}\frac{\partial^2U}{\partial e_{ij} \partial e_{kl}} + \frac{P}{2}(2\delta_{ij}\delta_{kl} - \delta_{il}\delta_{jk} - \delta_{ik}\delta_{jl}).
\label{cijkl}
\end{equation}
To investigate the potential application of the BK1 approach to the compressional stiffening
experiments reported here and in earlier experiments, we must consider the same deformations as in the experiments.
In doing so, we study both isotropic pre-stress and anisotropic
pre-stress. For the anisotropic pre-stress, we extend the BK1
approach. 

\subsection{Torsion with isotropic pre-stress}

\begin{figure}[htp]
\centering
  \includegraphics[width=0.48\textwidth]{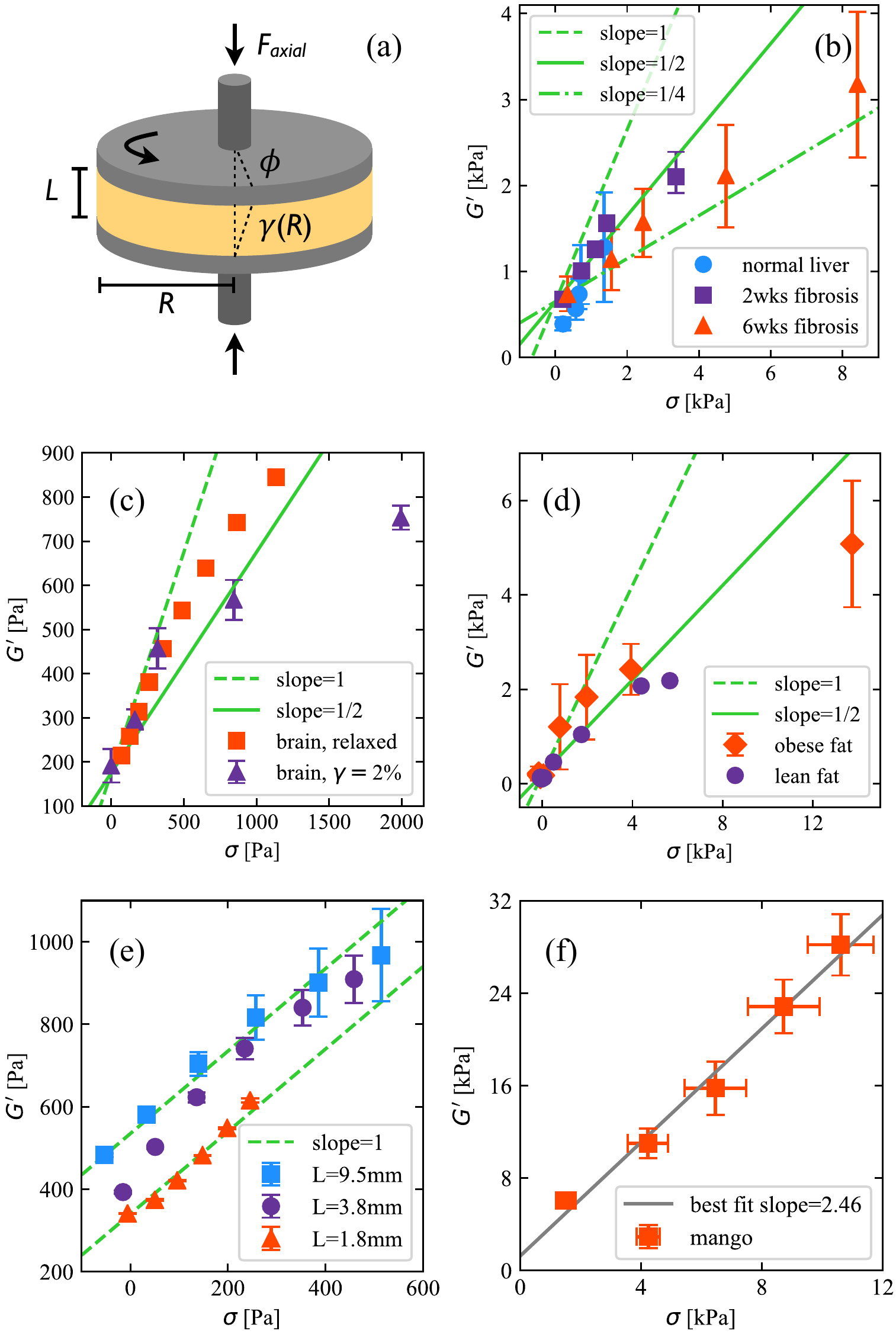}
  \caption{(a)~Schematic of rheometry experiment. In the limit $L\ll R$ and for adhesive boundary conditions, axial compression induces a volume change $\sim\pi R^2\Delta L$~\cite{Vahabi16, vanOosten16}. Outside this limit, the sample is free to bulge laterally, away from a cylindrical geometry~\cite{Perepelyuk16}, and any volume change associated with the compression is $\ll\pi R^2\Delta L$. Torsion measurements were carried out at fixed axial strain $\Delta L/L$, generating axial stress deviations $\delta\sigma$ from the zero torsion pre-stress $\sigma$, but these were typically small, i.e., $|\delta\sigma/\sigma|\ll 1$. (b)~Shear storage modulus $G'$ versus axial compressive pre-stress $\sigma$ for normal liver tissue and fibrotic liver tissue, two and six weeks after disease onset, replotted from Figure 5a of Perepelyuk, \emph{et al}.~\cite{Perepelyuk16}. The data points in each series correspond to axial strains of 0, 10, 15, 20, and 25\%. (c)~Compression stiffening data from normal mammalian brain tissue. Purple triangles are replotted from Figure 5c of Pogoda, \emph{et al}.~\cite{Pogoda14}. Axial strain ranges from zero to 40\%, in 5 and 10\% increments, respectively. (d)~Compression stiffening data from the fat tissue samples studied by Mihai, \emph{et al}.~\cite{Mihai15}. Axial strain again ranges from 0 to 40\%, in 10\% increments. (e)~Compression stiffening behavior of 2\% agarose gel. Different data series correspond to different sample aspect ratios (all have $R=1$ cm), with all three samples having been cut from the same master gel. Strain goes from 0 to 25\% in 5\% increments. The horizontal shifts required to collapse the $L$=9.5mm and $L$=3.8mm curves onto the $L$=1.8mm curve are a rough measure of the gravitational stresses present in the thicker samples -- see footnote above. (f)~Compression stiffening behavior of mango fruit flesh, with sample aspect ratio $L/2R=1/2$.   
}
  \label{janmey_data}
\end{figure}

Torsional deformation of a cylinder whose axis is situated at $x=y=0$ is equivalent to a symmetric strain
\begin{equation}
\left(
\begin{array}{ccc}
e_{xx} & e_{xy} & e_{xz}\\
e_{yx} & e_{yy} & e_{yz}\\
e_{zx} & e_{zy} & e_{zz}
\end{array}
\right)
=
\frac{\gamma(R)}{2R}\left(
\begin{array}{ccc}
0 & 0 & -y\\
0 & 0 & x\\
-y & x & 0
\end{array}
\right),
\label{eij}
\end{equation}
plus a rotation
\begin{equation}
\left(
\begin{array}{ccc}
w_{xx} & w_{xy} & w_{xz}\\
w_{yx} & w_{yy} & w_{yz}\\
w_{zx} & w_{zy} & w_{zz}
\end{array}
\right)
=
\frac{\gamma(R)}{2R}\left(
\begin{array}{ccc}
0 & -2z & -y\\
2z & 0 & x\\
y & -x & 0
\end{array}
\right).
\label{wij}
\end{equation}
Here $\gamma(r)=r\phi/L$ is the so-called ``torsional strain'' which has maximum value at $r=R$ (the cylinder radius), $L$ is the cylinder length, and $\phi$ is the angle of twist of one cylinder end with respect to the other (see Figure \ref{janmey_data}a). In order to apply the BK1 formalism, we decompose the solid cylinder into small volume elements, each of which experiences a local, homogeneous strain and undergoes a rigid rotation. Consider the element located at $(x_n=r_n,\,y_n=0,\,z_n)$ and having volume and energy $V_n$ and $U_n$, respectively. The only nonzero strain component is, switching to Voigt notation, $e_4=2e_{yz}=2e_{zy}=\gamma(r_n)$. Equation \ref{cijkl} then says that
\begin{equation}
c_{44} = \frac{1}{V_n}\frac{\partial^2U_n}{\partial [\gamma(r_n)]^2} - \frac{P}{2}.
\label{c44_isotropic}
\end{equation}
For an isotropic material with Lam{\'e} parameters $\lambda(=c_{12})$ and $\mu(=c_{44})$, this result extends to all volume elements, i.e., the shear modulus is given by
\begin{equation}
\mu = \frac{1}{V}\frac{\partial^2U}{\partial\gamma^2} - \frac{P}{2}.
\label{mu_isotropic}
\end{equation}

\subsection{Torsion with uniaxial pre-stress}
\label{uniaxial}

Next we consider the case $S_{ij}=-\sigma\delta_{iz}\delta_{jz}$. This form is perhaps easier to justify for the experiments in question, as no redistribution of the applied stresses, nor any volume change, is assumed. Unlike the isotropically pre-stressed reference state, however, the current one may have transversely isotropic material symmetry (assuming the initial, unstressed material was isotropic). In other words, the cells of the tissue may be flattened in the $z$-direction.

The counterpart to Equation \ref{cijkl} is straightforwardly obtained from BK1's Equation 4.21 for the strain energy density. We find
\begin{eqnarray}
c_{ijkl} = \frac{1}{V}\frac{\partial^2U}{\partial e_{ij} \partial e_{kl}} + \frac{\sigma}{4}&(&4\delta_{iz}\delta_{jz}\delta_{kl} - \delta_{iz}\delta_{kz}\delta_{jl} - \delta_{iz}\delta_{lz}\delta_{jk} \nonumber\\
&-& \delta_{jz}\delta_{kz}\delta_{il} - \delta_{jz}\delta_{lz}\delta_{ik}).
\label{cijkl_2}
\end{eqnarray}
Caution is needed here, because while $c_{ijkl}$ was guaranteed to be a well-defined elastic modulus tensor in the previous section, it is no longer so. There are two symmetry ``violations'' attributed to the uniaxial pre-stress, namely $c_{xxzz}-c_{zzxx}=-\sigma$ and $c_{yyzz}-c_{zzyy}=-\sigma$, according to BK1's Equation 4.20. Fortunately, $c_{44}$ is not directly implicated by these violations and presumably remains a valid elastic constant. Proceeding under this assumption, Equation \ref{cijkl_2} gives
\begin{equation}
c_{44} = \frac{1}{V}\frac{\partial^2U}{\partial \gamma^2} - \frac{\sigma}{4},
\label{mu_uniaxial}
\end{equation}
for an isotropic or transversely isotropic material under torsion.

\subsection{Apparent shear modulus}

Suppose one disregards the pre-stress and defines a quantity
\begin{equation}
G' \equiv \frac{1}{V}\frac{\partial^2U}{\partial \gamma^2} = \frac{\tau(R)}{\gamma(R)}.
\end{equation}
Here the latter equality involving the maximum shear stress $\tau(R)$
is obtained by integrating the energy density $\frac{\Delta U}{V} =
\frac{1}{2}G'\gamma^2(r)$ over the cylinder, and comparing the result
to Hooke's Law for torsion. Equation \ref{mu_uniaxial} now suggests
that a plot of $G'$ versus applied uniaxial pre-stress will be a
straight line having intercept $c_{44}$ and slope $1/4$, provided the
applied uniaxial pre-stress remains uniaxial within the sample. But if
boundary conditions, geometry, or some other factor  dictates that the
uniaxial pre-stress is redistributed into a hydrostatic pressure,
slope $1/2$ is predicted (by Equation
\ref{mu_isotropic}). Intermediate slope values are predicted in the
more general case where the applied compression generates a state of
pre-stress having both uniaxial and isotropic components. 

\subsection{Analysis}

In Figure \ref{janmey_data}, we compare this predicted range of slopes
(bracketed by solid and dash-dot lines) with compression stiffening
data from normal and fibrotic liver, brain, fat, agarose gel, and mango fruit flesh. With
no fitting parameters, we find good agreement with fibrotic liver and
reasonable agreement with lean and obese fat, but BK1 predicts too small
a rate of increase of $G'$ for the other biomaterials.  More
specifically, we find that 2 week old fibrotic liver redistributes its
internal stresses isotropically, while 6 week old fibrotic liver
redistributes its internal stresses tranversely to the parallel
plates. As fibrotic liver ages presumably there is more build up of
extracellular matrix (ECM) material intertwined within the cells. Perhaps such
fibrous material modulates how stresses are redistributed in the
presence of uniaxial compression. The normal liver data, on the other hand, does
not agree as well with either slope 1/4 or 1/2. 

Regarding fat tissue, there exists
better agreement with the
isotropically redistributed pre-stress calculation for lean fat
than for obese, though the obese data remain within one standard
deviation of slope 1/2 for all data points except the one at largest axial strain. Fat tissue is typically a tissue of high
expandability, however, in an obese state, adipocytes become
hypertrophic as a result of lipid uptake~\cite{Sun11}. Fat tissue also
contains ECM material. As fat tissue approaches the obese
state, there are changes in the ECM, mainly through an increasing deposition of collagen~\cite{Alkhouli13}.

The shear storage modulus versus axial stress for brain, agarose gel,
and mango fruit flesh data do not appear to exhibit the predicted BK1
behavior.  To explain these, we turn to a different theory.

\section{Birch approach and the role of the reference state}

\subsection{Background}
Several decades prior to BK1, Birch~\cite{Birch38} analyzed the case of
hydrostatic compression superposed with shear. He assumed that for an
applied stress of the form $T_{ij}=-P\delta_{ij} + T_{ij}'$, with
$T_{ij}'/P\ll1$, the strain response takes the form
$e_{ij}=\epsilon\delta_{ij} + e_{ij}'$, with
$e_{ij}'/\epsilon\ll1$. The quantity $T_{ij}'/e_{ij}'$ then defines a
modulus that is amenable to analytic calculation. While there are
several important differences between the Birch and BK1 treatments,
the one that places them on distinct conceptual footings is the
reference state, i.e. the configuration around which the energy
density is expanded. Birch's moduli are valid for a zero-stress
reference state, while BK1 addresses the pre-stressed reference state,
as detailed above. Another important difference is that finite strain
elasticity theory is invoked since the hydrostatic compression may be
of 
order of the moduli of the material~\cite{Birch38}. We will not here describe Birch's calculation in detail, because it is more complicated than BK1 and does not easily admit $P\delta_{ij}\to\sigma\delta_{iz}\delta_{jz}$ or other generalizations. Birch's result for the shear modulus of an isotropic material is
\begin{equation}
G = \mu + \frac{3(3-4\nu)}{2(1+\nu)}P,\label{BirchG}
\end{equation}
where $\nu$ is Poisson's ratio.  Unlike the BK1 approach, this latter
approach contains one fitting parameter in the form of Poisson's
ratio. Should the material be incompressible, then there is no fitting
parameter. It is illuminating that $\nu$ appears in the Birch approach, but does not appear in the BK1 approach, and this contrast is at the heart of the difference between the two approaches. In BK1, all of the compression that is going to happen has already happened (to the reference state). Therefore, the material's compressibility is irrelevant to any volume conserving deformations with respect to that reference state. In Birch, the shear response is coupled to compressibility insofar as both shear and compression are applied simultaneously to the reference state.

\subsection{Analysis} 
The Birch approach appears to describe those rheometry data in Figure
\ref{janmey_data} that are not well-described by BK1. In particular,
the variable Poisson's ratio can generate slopes $dG/dP$ ranging from
1 ($\nu=1/2$) to $9/2$ ($\nu=0$)~\footnote{It is worth noting that
  Mears, \emph{et al}. construct a shear modulus from Birch's result
  for Young's modulus, obtaining a result similar to, but not
  identical to, Equation~\ref{BirchG}. In their version, $dG/dP$
  ranges from 1 to 5~\cite{Mears69,JonesParry74}.}. With the exception
of liver and fat, the rheometry measurements reveal slopes
$dG'/d\sigma$ close to 1 for nearly incompressible ($\nu\approx1/2$)
materials such as brain tissue and agarose gel. Mango tissue, like other fruit, is more compressible due to the internal structure of
gas pockets~\cite{Ho09,Grotte02}. It exhibits a best fit slope of
$2.46$. This value is in remarkably good agreement with Equation
\ref{BirchG}, upon substituting Poisson's ratio of mango
($\nu=0.24\pm0.05$)~\cite{Jarimopas07}, which gives
$dG/dP=2.47\pm0.35$. That the Birch theory should work at all for the case of applied uniaxial stress, as opposed to applied isotropic stress, is perhaps surprising. Nevertheless, the excellent agreement of Equation \ref{BirchG} with mango, agarose gel, and (at small strains) brain tissue data, under $\sigma\to P$, suggests that $\sigma$ is internally redistributed into a hydrostatic pressure $P$, even outside the thin film limit discussed in the Introduction.

\section{Discussion}

We extend the list of biomaterials exhibiting compressional
stiffening to now include agarose gel and mango fruit flesh. The
ubiquitousness of compressional stiffening in tissues calls out for an
interpretative framework. Here, we provide that interpretative
framework via the application of two different classical elasticity results. The first is the BK1
approach, which has now been extended to the relevant experimental
geometry, and the second is the Birch approach. For both approaches the
shear storage modulus increases linearly with increasing axial stress and which
approach applies depends on the slope of the curve. The liver and fat
tissue exhibit BK1 behavior, while the agarose gel, brain, and mango
tissue exhibit Birch behavior. For those exhibiting the Birch 
behavior such that applied uniaxial stress is internally redistributed
as a hydrostatic pressure, these biomaterials appear to be behaving
qualitatively like an elastic bag filled with fluid.  Or to use a more
specific example, they behave like an automobile tire, which undergoes a small increase
in pressure when distorted from its optimal shape by a uniaxial
load. This picture is suitable for both animal and plant tissue,
despite the individual cell scales differences of cell walls,
vacuoles, and chloroplasts. At least in principle, it should be possible to directly test
for this stress redistribution by including a pressure gauge in the
rheometry apparatus. Importantly, such a redistribution does not imply
that the axial compression modulus observed in the rheometer
experiments should be similar to the tissue bulk modulus. That may be
the case in the thin film limit, but in general, the sample is free to
bulge out laterally during compression and any volume change is likely
$\ll\pi R^2\Delta L$.  

The necessity of the two approaches begs the question: how can we
predict which approach (i.e., reference state) is applicable for a particular sample?  It is interesting to note that the liver and fat tissue contain a fibrous protein extracellular matrix (ECM), while the brain
and mango tissue and agarose gel do not. Speculatively, the presence of this fibrous network could be 
the origin of the pre-stressed reference state required for BK1 behavior. Axial pre-stress might be generating contact changes in this network, and/or inducing anisotropies or other qualitative changes in the distribution of individual fiber tensions, such that the ``composite" material's response to subsequent shear stresses is altered from what it would be in the absence of a network component. In fact, Perepelyuk, \emph{et al}. argue that the interplay of an ECM component with a cellular component is what drives compression stiffening~\cite{Perepelyuk16}. Their proposed mechanism and conclusions are quite different than ours, as we will momentarily describe, but we do share a basic premise in that the fibrous ECM seems to play an important role in compression stiffening of liver tissue. In any case, a detailed understanding of BK1 and Birch's regimes of applicability to the parallel plate rheometry experiments would be an interesting direction for future work. 

Returning to the definition of $G'$ given above, it should be made
clear that this \emph{does} represent a kind of apparent mechanical
stiffness that may be said to exhibit compression stiffening. However,
any elastically isotropic material (according to BK1 theory) should
obey $G'(\sigma) = m\sigma + \mu$, with $1/4\leq m \leq 1/2$ depending
upon boundary conditions and the material's ability to redistribute
stresses, so biological tissues do not appear to be special in this
regard. What \emph{is} novel compressional stiffening behavior within
the context of parallel plate rheology, we suggest, are deviations
from this BK1 behavior, such as, perhaps, the Birch behavior which
\emph{requires} stress redistribution. Another type of deviation is
the compression softening behavior observed in the biopolymer network materials collagen and fibrin~\cite{Licup15, Vahabi16}. Intriguingly, these also stiffen in tension, with the same magnitude of slope ($dG/d|\sigma|=5$) as that appearing in Mears' version of the Birch theory for the appropriate Poisson's ratio ($\nu=0$)~\cite{Mears69,JonesParry74}.
Yet another kind of novelty would be a transition from one slope value to
another over time (given time-independent boundary conditions). As
mentioned earlier, the liver tissue data in Figure \ref{janmey_data}b hints at this. A possible interpretation is that some structural or compositional change occurs between two and six weeks after fibrosis onset that reduces the extent to which internal stresses are isotropically redistributed. 

One prior modeling effort to interpret the
observed compressional stiffening has already been mentioned. Perepelyuk, \emph{et al}. propose a
phenomenological model for simultaneous description of compression
stiffening, tension softening, and shear softening~\cite{Perepelyuk16}.
This model involves two components: an incompressible cellular phase
and a compressible filamentous (ECM) phase. Mechanical connections
between the two components are allowed to break under load, and
re-connect when the load is removed. Compression is thought to expel fluid through the
porous ECM phase, increasing the number of cell-cell contacts, and resulting in greater
resistance to shear. While reasonable agreement is
obtained with their liver data (replotted here in Figure
\ref{janmey_data}b), this agreement might be due to the fact that
there are at least five fitting parameters in the model (counting the
power law exponents.) Further, the reliance on two components is at
odds with agarose gel and with brain and mango tissue, the former lacking a cellular component and the latter lacking a filamentous component, but nonetheless exhibiting compression stiffening, qualitatively similar to that of liver tissue. Meanwhile, Mihai, \emph{et al}. address compression stiffening in homogeneous materials by showing that a subclass of Ogden hyperelastic models can account for compression stiffening in brain and fat tissue~\cite{Mihai15}, but again, these models have a large number of fitting parameters. In contrast, the BK1 and Birch theories provide a simple, universal explanation for compression stiffening and reasonably agree with available data spanning five different material types, the sole fit parameters being a binary choice of reference state (i.e., whether to apply BK1 or Birch), and in the case of Birch, the Poisson's ratio. For nearly incompressible materials, the latter fit parameter is effectively eliminated.

To further test the ideas herein against the models of Perepelyuk, \emph{et al}.~\cite{Perepelyuk16} and Mihai, \emph{et al}.~\cite{Mihai15}, we suggest that additional high-precision rheometer measurement be carried out for a variety of living and non-living soft materials, with simultaneous pressure measurement and supplementary Poisson's ratio measurement, if possible. Also, since in the BK1 theory it is $c_{44}$, not $G'$, that appears in the equation of motion and determines the speed of transverse sound $v_t = \sqrt{c_{44}/\rho}$, an independent measurement of sound velocity could constrain $c_{44}$ and verify that any pressure-dependence of $v_t$ enters only through the equation of state, $\rho(P)$. Finally, we mention that in the context of tumor identification and visualization, the distinction between $c_{44}$ and $G'$ should be important for certain types of ultrasound imaging, especially shear wave elastography~\cite{Sarvazyan98}.

\begin{acknowledgments}
TAE wishes to thank Daniel Sussman for providing some useful references. TAE and JMS acknowledge financial support from NSF-DMR-CMMT Award Number 1507938. KP, KC, and  PAJ acknowledge financial support from NSF 16-545  DMR 17-20530 and NIH U54-CA193417.
\end{acknowledgments}

%

\end{document}